\documentstyle[12pt,aaspp4]{article}
\def\ls{{_<\atop^{\sim}}}

\def\masomen{{_+\atop^{-}}}

\begin{document}

\title{The Equilibrium Photoionized Absorber in 3C351}

\author{ Fabrizio Nicastro$^{1,2}$, Fabrizio Fiore$^{1,2,3}$, 
G. Cesare Perola$^4$, Martin Elvis$^1$}

\affil {$^1$ Harvard-Smithsonian Center for Astrophysics\\ 
60 Garden St, Cambridge MA 02138}

\affil {$^2$ Osservatorio Astronomico di Roma\\
via Osservatorio, Monteporzio-Catone (RM), I00040 Italy}

\affil {$^3$ SAX Science Data Center\\
via Corcolle 19, Roma I00100 Italy}

\affil {$^4$ Dipartimento di Fisica, Universit\`a  degli studi
``Roma Tre''\\ 
Via della Vasca Navale 84, Roma, I00146 Italy}

\author{\tt version: 18:00, 8 August, 1998}

\begin {abstract}

We present two ROSAT PSPC observations of the radio-loud,
lobe-dominated quasar 3C~351, which shows an `ionized absorber' in
its X-ray spectrum. The factor 1.7 change in flux in the
$\sim$2~years between the observations allows a test of models
for this ionized absorber.

The absorption feature at $\sim0.7$ keV (quasar frame) is present
in both spectra but with a lower optical depth when the source
intensity - and hence the ionizing flux at the absorber - is higher,
in accordance with a simple, single-zone, equilibrium
photoionization model. Detailed modeling confirms this agrement
quantitatively. The maximum response time of 2~years allows us to
limit the gas density: $n_e > 2 \times 10^4$ cm$^{-3}$; and the
distance of the ionized gas from the central source R $<$ 19
pc. This produces a strong test for a photoionized absorber in
3C~351: a factor 2 flux change in $\sim$1 week in this source
{\em must} show non-equilibrium effects in the ionized absorber.

\end{abstract}

\newpage
\section{\bf INTRODUCTION}

The most luminous known AGN with an ionized absorber is the radio-loud 
lobe-dominated
\footnote{only 0.65 \% of the flux density is contained in the compact 
core at 6 cm, (Kellerman et al., 1989).}
quasar 3C351 (L$_{0.1-2 keV} = 2.3\times10^{45}$ erg s$^{-1}$, z=0.371, 
Fiore et al., 1993). 
Ionized absorbers are common in low redshift, low luminosity Seyfert 
1 galaxies (Reynolds, 1997), but rare in higher redshift, higher 
luminosity quasars. High luminosity AGNs are very likely physically 
larger and so may exhibit slower time variability and have different 
physical conditions. 
3C~351 has a very high UV to X-ray ratio ($\alpha_{OX}=1.55$, Tanambaun 
et al., 1989) and its IR to UV spectrum does not show any evidence of 
reddening, in contrast to several low redshift Seyfert 1 galaxies 
thought to host dusty/warm absorbers
\footnote{i.e. IRAS~13349+2438, Brandt et al., 1996, 1997; 
IRAS~17020+4544, Leighly et al., 1997, Komossa \& Bade, 1998; 
MCG-6-30-15, Reynolds et al., 1997.} 
. Only two other radio-loud quasars, 3C~212 (Mathur et al., 1994) and 
3C~273 (Grandi et al, 1997), have candidate ionized absorbers. 

The main absorption feature in the ROSAT PSPC X-ray spectrum of 
3C~351 is a deep edge at $\sim0.7$ keV (quasar frame), which is likely 
to be due to OVII-OVIII. Mathur et al. (1994) built a simple one-zone 
model of this absorber that also explained the OVI, CIV and Ly$\alpha$ 
UV absorption lines seen in a nearly simultaneous HST FOS spectrum. 
Such simple models, although elegant, have come under criticism. 
Ionized absorbers in some Seyfert galaxies show different variability
behavior in different absorption features, some of which are not
as predicted for gas in photoionization equilibrium. Multiple
absorbing zones have been introduced to explain these effects
(e.g. Reynolds 1997). In another paper (Nicastro et al.,
1998) we instead explore additional physics (non-equilibrium 
photonionization models and collisional models) while retaining a 
single zone of absorbing gas. 
3C~351 is a relatively slowly varying source, compared to the rapidly 
variable Seyfert with warm absorbers. 
This persistent variability on timescales that span possible ionization 
and recombination times makes it hard to define an initial equilibrium 
state. In contrast 3C~351 presents a simplified situation in which 
to study changes in the ionization state in response to luminosity 
changes. 
Here we show that in 3C~351 at least the simplest photoionization 
equilibrium, single zone model continues to be sufficient.

3C351 was observed twice by ROSAT, on 1991 October and 1993
August.  The first observation was reported by Fiore et
al. (1993). The second observation was then proposed in order to
search for time variability that could test photoionization
models. Fortunately a factor 1.7 decrease in the PSPC count rate
was seen, providing just such a test. Here we present the second
data set and compare the results with predictions. 

\section{Observations and Data Analysis}

We considered the two PSPC observations of 3C351 taken roughly
two years apart, in 1991 October and 1993 August. Table 1 gives,
for each observation, the observation date, its duration, the net
exposure time, the net source counts, the count rate, and the
signal to noise ratio.  The data reduction and the timing
analysis were performed using XSELECT. 

We reduced the second PSPC observation of 3C351 following the
procedures used in Fiore et al. (1993) for the first observation.
For the first observation we used the spectrum and light curve
obtained by Fiore et al. (1993). The source intensity was
consistent with a constant value during both PSPC observations, 
with no more than 20\% change in the $\sim 2$ day and $\sim 13$ 
day spans of the two observations. 
3C~351 thus appears more stable than the lower luminosity seyferts 
with ionized absorber (Reynolds, 1997). Since 3C~351 has a 
flat X-ray spectrum and broad optical emission lines it is consistent 
with the trend exhibited by the three similar but radio-quiet, 
PG quasars of the Fiore et al. (1998) sample. 
Those quasars show little or no variability on timescales as short as 
10 days (compared to the narrow line objects of that sample which are 
rapidly variable and have steep X-ray spectra).
However the mean count rate dropped by a factor $\sim1.7$ during the
22 months between the observations (see Table 1), again consistently 
with the Fiore et al. (1998) discovery. 
Hereinafter we shall call these the 'High' and 'Low' states,
respectively.

\footnotesize
\begin{table*}
\begin{center}
\caption{\bf\small ROSAT PSPC observations of 3C~351}
\vspace{0.4truecm}
\begin{tabular}{|cccccccc|}
\hline
Start Date & Duration & Exposure & Net Counts & Count Rate & S/N
& State & ROR \\
           & (ksec)   & (ksec)   &            & ct~s$^{-1}$&  & \\
\hline
91 Oct 28 &  168 & 13.1 & 1390 & 0.110 & 31.0 & High & rp700439 \\
93 Aug 23 & 1143 & 15.4 &  980 & 0.064 & 26.0 & Low  & rp701439 \\
\hline
\end{tabular}
\end{center}
\end{table*}
\normalsize
\section{Ionization Models}

We produced equilibrium photoionization models and pure
collisional ionization models using CLOUDY (version 90.01,
Ferland 1996) to fit to the PSPC spectra of 3C~351. We use a
Friedman cosmology with $H_0=50$ km s$^{-1}$ Mpc$^{-1}$ and
$q_0=0.1$ to derive a luminosity of $L_{2keV}$ $\sim 3\times
10^{44}$ erg s$^{-1}$ kev$^{-1}$ for the quasar. This determines the
distance scale in the absorber for a given value of the
ionization parameter, U.

3C~351 is a radio-loud lobe-dominated quasar and, as Fiore et al
(1993) and Mathur et al (1994) have shown, the ionization state
of the gas strongly depends on the spectral energy distribution
(SED) of the ionizing continuum, from radio to hard X-rays.  We
therefore made our photoionization models using the observed SEDs
for 3C351 (Mathur et al 1994).  We varied the ill-determined
energy at which the X-ray spectrum turns up to meet the
ultraviolet using a broken power-law with a break energy from the
unobserved EUV range to well within or above the PSPC range, as
in Mathur et al (1994).  
The low energy spectral index is fixed by the observed flux in the 
UV at one end, and the break energy at the other. 
The high energy spectral index above the break energy is fixed at 
$\alpha_E=0.9$. This is typical for lobe dominated radio loud quasars 
(Shastri et al., 1993). 
In the following we present a spectral analysis using a photoionizing 
continuum with a break energy of 0.4 keV (quasar frame) and UV-to-soft-X-ray 
index of $\alpha_E=1.5$. 
We note however that this continuum model, although providing 
a good description of the PSPC data, should only be thought as one 
of the possible continuum parameterization connecting the UV and the soft 
X-ray data of 3C~351. No break energy is required by data, which are 
well fitted by a simple power law with $\alpha_E=0.9$ across the whole 
PSPC band.

\subsection{Spectral Analysis}

The spectral analysis was performed using XSPEC and the latest
version of the 256 channels PSPC response matrix (``pspcb\_gain2\_256.rsp'', 
released on January 12, 1993). 
We binned each spectrum in channels with at least 50 counts, to warrant  
the poissonian statistics to be used. 

We first fitted the high and low state spectra with a simple
power law model reduced at low energy by Galactic absorption
($2.26 \times 10^{20}$ cm$^{-2}$, Elvis et al., 1989). 
Only 2 parameters were left free to vary: the photon
spectral index $\Gamma$ and the normalization $F_0$.  
In both the cases the $\chi^2$ is unacceptably high: $\chi^2_r = 2.99$ 
and 3.09, for 46 and 19 degree of freedom respectively.  The upper 
panels of Fig. 1 (a) and (b), show the data and the best fit simple power 
law models folded with the PSPC response, for the High and Low state 
respectively ($\alpha_E^{High}= 1.3$ and $\alpha_E^{Low}=1.0$). 
Lower panels show the ratio between the data and the best fit models. 
The deficit of counts at $\sim 0.6$ keV (rest frame) is evident in both 
spectra. 
We tested more complex models for the intrinsic continuum, 
adding a second power law, and a low energy black body component. In the 
latter case the $\chi^2$ was still unacceptably high, while a double 
power law, though producing acceptable fits by the $\chi^2$ point of 
view, gave uncommon spectral index values for this class of sources. 
Furthermore in both the cases negative residuals were still evident 
around 0.6 keV (observed). 
Following Fiore et al. (1993) we then interpret these features as due 
to a blend of OVII-OVIII absorption K-edges caused by ionized gas
along the line of sight. A test of this by making a fit of a single power 
law with a notch at 0.6 keV (observed) -- a simplistic ionized absorber 
model -- gave acceptable $\chi^2$: 1.1 (45 dof) and 1.0 (18 dof) for the 
High and the Low state respectively. 

\begin{figure}
\epsfysize=3.2in 
\epsfxsize=3in 
\hspace{1cm}\epsfbox{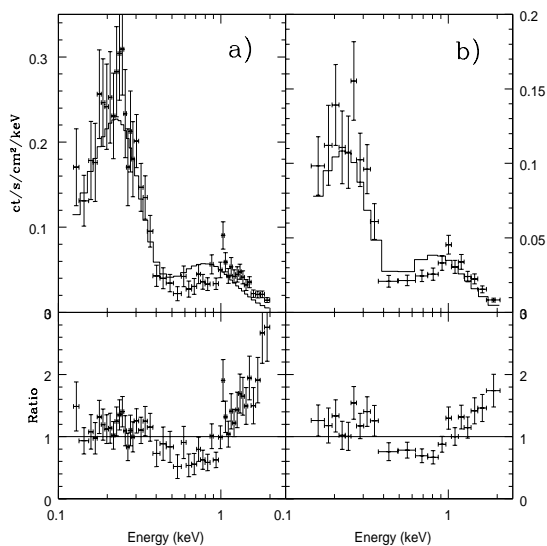} 
\vspace{0in}\caption[h]{\footnotesize High and low state PSPC spectra of 
3C~351 along with the best fit simple power law models folded with the PSPC 
response (Fig. 1a and 1b respectively, upper panels). Ratio between the data 
and the best fit models (lower panels).}
\end{figure}

We then fitted both spectra using the equilibrium photoionization
models described above.  Only three parameters were
left free to vary in the fit: the normalization ($F_0$), 
the ionization parameter U$=\left( \int_{1Ryd}^{+\infty} F_E dE \right) / 
(4 \pi R^2 n_e c)$  and the warm absorber column density $N_H$. 
The soft (E$ < 0.29$ keV) and hard ($0.29 <$ E $< 2.5$ keV) spectral index 
were fixed at $\alpha_E^{UV-X}=1.4$ and $\alpha_E=0.9$, the values
of the ionizing continuum used to build the models (see above). 
Table 2 gives the best fit parameters.  While the best fit values
of $N_H$ are consistent with a constant value, there is 
marginal evidence (at the 2 $\sigma$ level) of variation in U.

We then fitted the high and low state spectra simultaneously,
fixing $N_H$ to the best fit value found in the high state, since
this has the best statistics. Table 2 gives again the best fit
values for U and the $F_0$ in the two spectra (with the 90\%
confidence intervals for one interesting parameter).  The best
fit U and $F_0$ are linearly correlated with each other (Figure
2), in the manner expected if the gas is in photoionization
equilibrium with the ionizing continuum .  Table 2 also gives the
best fit abundances of OVII and OVIII and their ratio. The change
of U translates to a factor of three change in the OVII to OVIII
ratio as the source intensity drops by about 70 \%.

\footnotesize
\begin{table*}
\begin{center}
\caption{\bf Equilibrium Photoionization Model Fits}
\begin{tabular}{|cccccccc|}
\hline
Spectrum & Log($N_H$) & Log(U) & $^aF_0$ & $\chi^2$(d.o.f.) & $n_{OVIII}$ & 
$n_{OVIII}$ & $n_{OVIII}/n_{OVII}$ \\
\hline
High & $22.14_{-0.17}^{+0.14}$ & $0.78_{-0.13}^{+0.21}$ & $8.9\masomen1.8$ 
& 0.89(45) & 0.783 & 0.168 & 0.03 \\
Low &  $21.89_{-0.28}^{+0.22}$ & $0.49_{-0.30}^{+0.23}$ & $5.0\masomen0.1$ 
& 0.89(18) & 0.822 & 0.081 & 0.10 \\
High+Low & 22.14 (frozen) & $0.78_{-0.09}^{+0.12}, 0.63_{-0.09}^{+0.10}$ & 
$8.7_{-0.8}^{+0.9}, 6.0_{-0.5}^{+0.7}$ & 0.99(62) & & & \\
\hline
\end{tabular}
\end{center}
$^a$ in $10^{-5}$ ph s$^{-1}$ cm$^{-2}$ keV$^{-1}$ (at 1 keV). 
\end{table*}
\normalsize

\bigskip

\begin{figure}
\epsfysize=4.5in 
\epsfxsize=3.in 
\hspace{1cm}\epsfbox{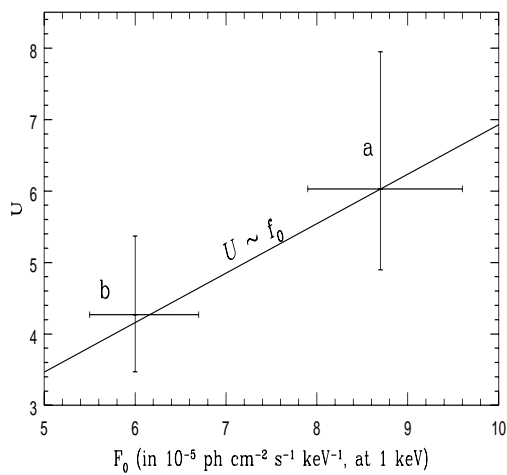} 
\vspace{-1.in}\caption[h]{\footnotesize Best fit ionization
parameter U versus best fit 1 keV normalizations $F_0$ for
spectra 'High' and 'Low'. The solid line shows the linear relationship
U($F_0$) expected if photoionization equilibrium applies.
}
\end{figure}

Collisional ionization models can fit the data equally well, as they
should at low column densities (Nicastro et al., 1997).  To do so
however they require arbitrary changes, by a factor 1.5 in temperature or
a factor 2 in N$_H$. Instead photoionization equilibrium models
predict the observed correlation with the ionizing continuum.
To the authors this is a strong argument in favor of
a photoionization model. 

%
%
%
%
%

\subsection{X-ray Colors of the warm absorbers in 3C~351}


The behavior of the main physical properties of the absorber can
be seen in a color-color diagram.  In Fig. 3 we plot the hardness
ratios HR=H/M against the softness ratio SR=S/M from the count
rates in three bands (S=0.15-0.58~keV, M=0.88-1.47~keV, and
H=1.69-3.40~keV, at z=0.371) for theoretical curves (for
log($N_H$)=21, 21.5, 22 and 22.14) obtained by folding the
equilibrium photoionization models (for log(U) in the range
-0.3-1.5, and Galactic $N_H$) with the response matrix of the
PSPC.

\begin{figure}
\epsfysize=4.0in 
\epsfxsize=4.0in 
\hspace{1cm}\epsfbox{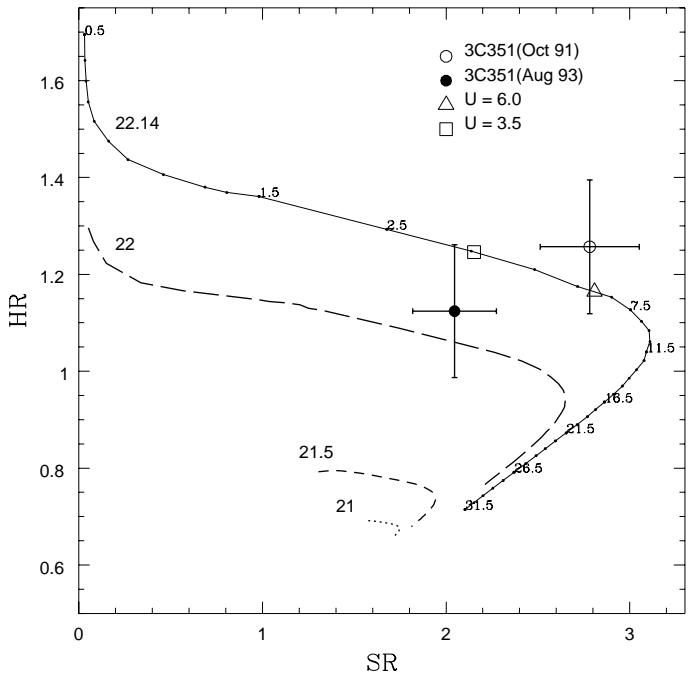} 
\vspace{0in}\caption[h]{\footnotesize 
Color-color diagram of the two observations of 3C351. Lines are
theoretical photoionization curves, built folding equilibrium
photoionization models with the response matrix of the
PSPC. Different lines correspond to two different values of the
gas column density: log($N_H$)=22.14 (solid line), log($N_H$)=22
(dashed line). On each curve the ionization parameter U increases
going from top to bottom. U values are indicated on the
log($N_H$)=22.14 line.  On the same curve we mark two points: a)
the best fit U to the october 1991 observation, and b) this value
scaled by the factor 1.7 (the ratio of the source fluxes).}
\end{figure}

The position of a point in this diagram readily gives the
dominant ion in the gas.  The rapid change of SR as U increase
from 0.5 to 5.5 (on the curve corresponding to log($N_H$)=22.14)
corresponds to large increases of the transparency of the gas at
E$<0.5$ keV as H and He becomes rapidly fully ionized. As U
further increases from 5.5 to 16.5 the SR color changes more
slowly, and inverts its trend at U$\sim 11$, with HR now changing
more rapidly than SR.  The inversion point indicates the switch
from an ionization state dominated by OVII to that dominated by
OIX.  In the last part of the curve as U increases, SR and HR
decrease until all the ions in the gas are fully stripped and the
gas is completely transparent to radiation of any energy.

The two data points show the two 3C~351 observations.  The best
fit U in the High observation is marked on the log($N_H$)=22.14
curve along with the value obtained by scaling by the intensity
ratio between the two observations (a factor 1.7).  The two
points are consistent with the position of the observed colors of
3C351 in the two observations, so the predictions of the simple
equilibrium photoionization model is consistent with the data.


\section{Discussion}

The as predicted change of the ionization parameter in the
3C~351 absorber to a change in the ionizing continuum is strong
evidence that photoionization is the dominant ionization
mechanism.  We also see that the absorber comes to ionization
equilibrium within 22 months, and can use this to constrain the
physical properties of the absorber.


The time $t_{eq}$ measures the time the gas needs to reach
equilibrium with the instantaneous ionizing flux (Nicastro et al,
1997). This time depends on the particular ionic species
considered.  The ionic abundances of Oxygen in the absorber in
3C~351 are distributed mainly between only two ionic species:
OVII and OVIII. In this simple case a useful analytical
approximation for $t_{eq}$ is:

\begin{eqnarray}
\small
t_{eq}^{OVII,OVIII}(t\to t+\Delta t) \sim 
{1 \over {\alpha_{rec}(OVII,T_e)_{eq} n_e}} 
{1 \over {\left[ {\left( \alpha_{rec}(OVI,T_e) \over 
\alpha_{rec}(OVII,T_e) \right)_{eq}} + {\left( {n_{OVIII} \over {n_{OVII}}} 
\right)_{eq}} \right]_{t+\Delta t}}}\nonumber
\normalsize
\end{eqnarray}

where {\em eq} indicates the equilibrium quantities. 

By requiring that $t_{eq}$ for OVII and OVIII species is shorter
than the $4.2\times10^7$ s (quasar frame) elapsed between the two 
observations, we can find a lower limit to the electron density of the 
absorber: $n_e> 2.5 \times 10^3$ cm$^{-3}$. Since we know the values of U
and $F_0$, this density limit translates into a limit on the
distance of the ionized gas from the central source, R$<50$ pc.

These limits are consistent with the upper limit of $n_e<5\times
10^7$ cm$^{-3}$, and R$>0.3$ pc, found by Mathur et al. (1994)
using a lower limit for the distance of the cloud from the
central source, based on the absorber being outside the broad
emission line region. 

The relative closeness of these two limits (factor $\sim$100)
implies that a variation in shorter times ($\sim 1$ month), would
be likely to show non-equilibrium effects (Nicastro et al, 1997).
If a factor 2 flux change in $\ls 1$ week showed no such effects
the simplest photoionization model would have to be abandoned.

\section{Conclusion} 

We have tested ionization models for the ionized absorber in
3C~351. In particular, we tested a simple one zone
photoionization equilibrium model on two PSPC spectra of 3C~351
that show a factor $\sim2$ decrease in flux.  The model correctly
predicts the sense and amplitude of the observed change in the
ionization state of the absorber, correlated with the ionizing
continuum flux.

Given that photoionization equilibrium applies we can derive a
lower limit to the electron density of the absorber:
$n_e>2.5\times 10^3$ cm$^{-3}$.  This is consistent with the
upper limit of $n_e<5\times 10^7$ cm$^{-3}$ found by Mathur et
al. (1994).  The distance of the ionized gas from the central
source is then 0.3 pc $<$ R $<$ 50 pc.  The closeness of these
two limits creates a strong test of photoionization models:
factor 2 variations in 3C~351 on timescales of order a week {\em
must} show non-equilibrium effects.

\bigskip
\center{\bf Aknowledgements}

This work was supported in part by NASA grants 
NAG5-3066 (ADP),  NAG5-2640 (ROSAT), NAGW-2201 (LTSA)




\begin{references}

Brandt W.N., Fabian A.C., Pounds K.A., 1996, MNRAS, 278, 326

Brandt W.N., Mathur S., Reynolds C.S., Elvis M., 1997, MNRAS, 292, 407

Elvis M., Wilkes B.J., Lockman F.J., 1989, A.J., 97, 777  

Ferland G.J., 1996, CLOUDY: 9001

Fiore F., Elvis M., Mathur S., Wilkes B.J., McDowell J.C. 1993, ApJ, 415, 192

Grandi P., et al., 1997, A\&A, 325, 17

Kellerman K.I., Sramek R., Schmidt M., Shaffer D.B., \& Green R., 1989, 
AJ, 98, 1195

Komossa S. \& Bade N., 1998, A\&A, 331, 49

Leighly K.M., Kay L.E., Wills B.J., Wills D., \& Grupe D., 1997, ApJL, 
489, 137

Mathur S., 1994, ApJ, 431, 75

Mathur S., Wilkes B.J., Elvis M., Fiore F. 1994, ApJ, 434, 493

Nicastro F., Fiore F., Perola G.C., Elvis M., 1998, Apj submitted

Reynolds C.S., 1997, MNRAS, 286, 513

Reynolds C.S., Ward M.J., Fabian A.C. \& Celotti A., 1997, MNRAS, 291, 403

Shastri P., Wilkes B.J., Elvis M., McDowell J. 1993, ApJ, 410, 29

Tanambaum H., Avni Y., Green R.F., Schmidt M., \& Zamorani G., 1986, ApJ, 
305, 57

\end{references}
\end{document}